# Annealing-Induced Structural and Optical Evolution of Ion-Beam-Sputtered $Ta_2O_5$ Films for High-Power Laser Applications

**Chang Chen[a,b,c,†], Yuan Fang[a,b,†], Yanzhi Wang[a,b,*], Jijun Feng[c], Yesheng Lu[a,b,*], Yu Chen[a,b], Ge Zhang[a,b], Jianda Shao[a,b,d,e,*]**

[a] *Laboratory of Thin Film Optics, Shanghai Institute of Optics and Fine Mechanics, No. 390 Qinghe Road, Jiading District, Shanghai 201800, China*

[b] *Key Laboratory of Materials for High Power Laser, Shanghai Institute of Optics and Fine Mechanics, No. 390 Qinghe Road, Jiading District, Shanghai 201800, China*

[c] *Shanghai Key Laboratory of Modern Optical System, Engineering Research Center of Optical Instrument and System (Ministry of Education), School of Optical-Electrical and Computer Engineering, University of Shanghai for Science and Technology, Shanghai 200093, China*

[d] *CAS Center for Excellence in Ultra-intense Laser Science, Shanghai 201800, China*

[e] *Hangzhou institute for advanced study, Hangzhou 310024, China*

[†]These authors contribute equally.

**yanzhiwang@siom.ac.cn*

**luyesheng@siom.ac.cn*

**jdshao@siom.ac.cn*

**Abstract:** Tantalum pentoxide ($Ta_2O_5$) thin films deposited by ion beam sputtering (IBS) are essential components in high-power laser systems, where post-deposition annealing is routinely used to optimize film quality. In this work, we systematically investigate the structural and optical evolution of IBS-deposited $Ta_2O_5$ single layers subjected to post-deposition annealing at temperatures from 400 to 800 °C. The films undergo an amorphous-to-polycrystalline transition between 600 °C and 700 °C, accompanied by a marked increase in surface roughness. Argon bubbles inherently incorporated during IBS serve as structural markers: their size correlates positively with the surface roughness from the as-deposited to 700 °C but decouples at 800 °C, revealing a transition from matrix-relaxation-dominated to surface-diffusion-dominated behavior. Transmittance spectra show a red shift at 400 and 600 °C which can be explained by physical thickness increase. Weak absorption decreases from ~6 ppm in as-deposited films to ~2 ppm at 400–600 °C due to annihilation of oxygen vacancies, rises markedly at 700 °C due to grain-boundary defects and then drops again at 800 °C as grain boundaries relax. We identify 400 °C as the optimal annealing window for IBS-deposited $Ta_2O_5$ single layers, with the lowest surface roughness and effectively reducing weak absorption while preserving the amorphous phase.

**Keywords**: $Ta_2O_5$; Ion beam sputtering; Annealing; Weak absorption; Argon bubbles

## I. Introduction

Tantalum pentoxide ($Ta_2O_5$) thin films have attracted extensive attention in high-power laser systems and gravitational wave detectors due to their high refractive index, broad transparency window, excellent thermal stability, and high laser-induced damage threshold[1–6]. Among various deposition techniques, ion beam sputtering (IBS) is widely adopted for producing dense, smooth, and highly reproducible oxide coatings[7,8]. However, a well-recognized drawback of IBS is the unintentional incorporation of the working gas (argon) into the growing film. Energetic Ar ions or neutrals may be backscattered from the target and subsequently trapped in the film matrix[9–12]. Post-deposition annealing is routinely employed to improve film stoichiometry, reduce oxygen vacancies, and thereby lower weak absorption[13,14]. Nevertheless, annealing can also trigger undesirable effects, such as amorphous-to-polycrystalline phase transition, stress relaxation, and microstructural degradation[15,16]. Considering the demand of improving film quality, especially minimizing optical loss in high-power laser applications, it is important to understand how annealing affects structural and optical properties of $Ta_2O_5$ films with the existence of Ar bubbles.

Paolone et al. have recently documented structural and optical changes in IBS $Ta_2O_5$ upon thermal treatments[11]. They found a release of argon in concomitance with crystallization and a decrease of the oxygen stoichiometry for samples heated in vacuum. However, considering the real preparation conditions of mirror coatings used for high power laser systems, the systematic investigation of annealing in air is more demanding. Peters et al. have found that the high pressure inside the IBS-induced nanobubbles in hafnia films reduces the threshold for laser-induced plasma breakdown leading to film damage[17]. However, different from nonlinear multiphoton absorption, how Ar bubbles affect weak absorption of IBS $Ta_2O_5$ films remains an open question.

In this work, we reported comprehensive investigation of IBS $Ta_2O_5$ single layers annealed at temperatures ranging from 400 to 800 °C. Based on our structural, morphological and optical results, we classified the annealing behavior of the films into four distinct stages. We identified 400 °C as the optimal annealing temperature and 400–600 °C as an effective low-absorption window. We found below 700 °C Ar bubbles can serve as structural marker while at 800 °C they decouple. Through tracking the evolution of Ar bubbles and weak absorption with annealing simultaneously, we established that linear optical loss is governed by electronic defects rather than argon-filled cavities.

## II. Methods

The samples were prepared by the double ion beam sputtering deposition device of Veeco Spector. The main ion source working gas is argon which is used to sputter the target material. The targets are metal Ta and $SiO_2$ with purity greater than 99.99%. The vacuum degree of the chamber is about $10^{-4}$ Pa. The base baking temperature is 90 °C. The substrates are fused silica glass and single-crystal silicon. The thickness of the prepared $Ta_2O_5$ and $SiO_2$ monolayer films is about 1000 nm. $Ta_2O_5$ samples were annealed in air at 400, 600, 700, and 800 °C for 5 hours.

The cross-section morphology of the samples was observed by STEM (Thermo Fisher Company) under a 200 kV acceleration voltage. STEM can also provide energy-dispersive X-ray spectroscopy (EDS) for element scanning analysis[18]. In high-angle annular dark-field (HAADF) imaging mode, Z-contrast arises from density differences between the bubble and the material surrounding it[19]. Here, the argon content in $SiO_2$ and $Ta_2O_5$ films were studied by combining HAADF imaging and EDS. The collected section morphology was analyzed using STEM/EDS elemental microanalysis software. X-ray diffraction (XRD, Empyream, PANalytical) was used to measure the crystallization of samples after annealing. The measurement angle was 10°-90°, and the minimum step size was 0.0001°. The root-mean-square (RMS) roughness of the sample surface was evaluated using atomic force microscopy (AFM, Dimension-3100 Veeco). Five points were tested for each sample (within 0.01 nm error) to obtain an average value. The transmittance of the as-deposited and the annealed samples was measured by Lambda1050 dual-channel spectrophotometer with a measurement accuracy of ±0.08%. The weak absorption of the samples at 1064 nm was measured according to the principle of surface thermal lens technology[20].

## III. Results and discussion

### A. Microstructure and argon incorporation in $Ta_2O_5$ films

Fig.1(a)–(d) and (e)–(h) show the cross-section morphology and EDS element scanning results of $SiO_2$ and $Ta_2O_5$ single layers, respectively. As shown in Fig.1(a), no obvious defects

such as holes are observed in the $SiO_2$ film. The scanning result of the Ar element in Fig. 1 (b) indicates that the Ar content is relatively small, and no aggregation is observed. While as shown in Fig. 1 (e), many holes can be observed in the cross-section morphology of $Ta_2O_5$ film. Compared with $SiO_2$ film, the relative content of the Ar element as shown in in Fig. 1 (f) increases, and the Ar element is aggregated. As can be seen by comparison of red circled area in Fig.1 (e) and (f) and by EDS line scanning element content analysis shown in Supplementary Material Fig. S1, the Ar aggregation area corresponds to the hole positions. Our results are consistent with results reported previously. In titanium dioxide doped $Ta_2O_5$ thin films, researchers have found that these Ar-aggregated holes are actually argon bubbles[9]. In IBS produced hafnia films, researcher have found that the replacement of the Ar working gas by Xe reduces the gas incorporation in the films from ~6% for Ar to less than 1.8% for Xe. They attribute this difference to the amount of energy transferred[10]. In our case, the Ar content is much less in $SiO_2$ films than in $Ta_2O_5$ films. This is because the mass ratio between Ar and Si is much closer to 1 than that between Ar and Ta, and thus the Ar ions lose more energy and are less incorporated in the $SiO_2$ films.

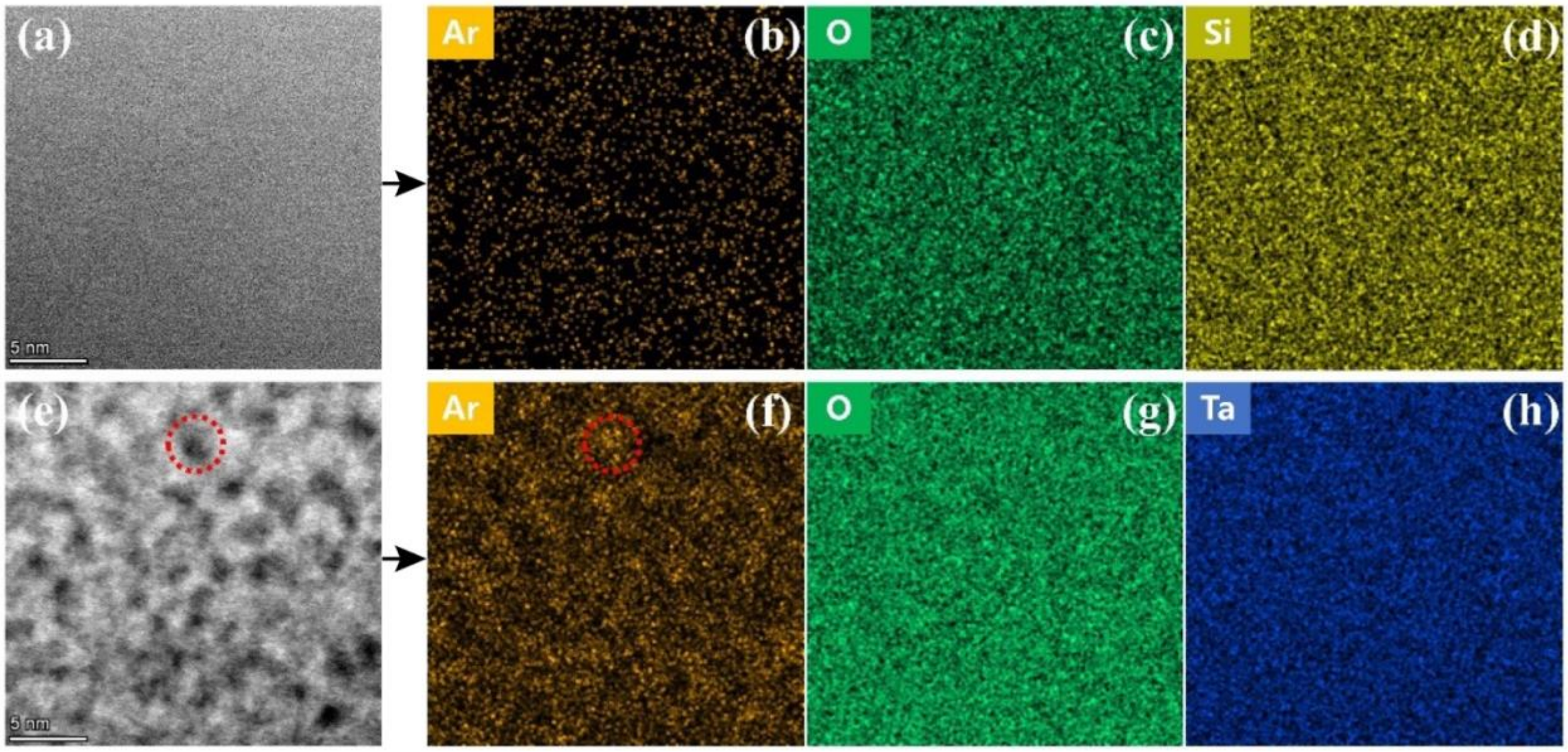

Fig. 1. Cross-section morphology observed under STEM/HAADF and EDS element scanning. (a) $SiO_2$ single layer cross-section morphology; (b), (c), and (d) show EDS element scanning of Ar, O, and Si in $SiO_2$ film, respectively; (e) $Ta_2O_5$ single layer cross-section morphology; (f), (g), and (h) show EDS element scanning of Ar, O, and Ta in $Ta_2O_5$ film, respectively.

**B. Annealing-induced structural and morphological evolution**

The $Ta_2O_5$ single layer samples prepared simultaneously were annealed to study the effect of annealing temperature on film properties. We first performed XRD measurements on samples with different annealing temperatures. As shown in Fig. 2, when annealed below 600 °C the XRD curves contain broad valleys and humps which indicates that the samples are in the amorphous phase, while multiple sharp diffraction peaks appear when annealing at 700 and 800 °C which indicates that polycrystalline phases are formed at these temperatures. The diffraction peak intensities at 700 and 800 °C are reduced five-fold for clear illustration. Note that the diffraction peak intensity at 700 °C is significantly higher than that at 800 °C. This can be explained by the increased lattice disorder induced by larger bubbles (Fig. 3e) and the smaller crystallite sizes, as evidenced by the larger FWHM of XRD peaks at 800 °C (extracted from Fig. 2).

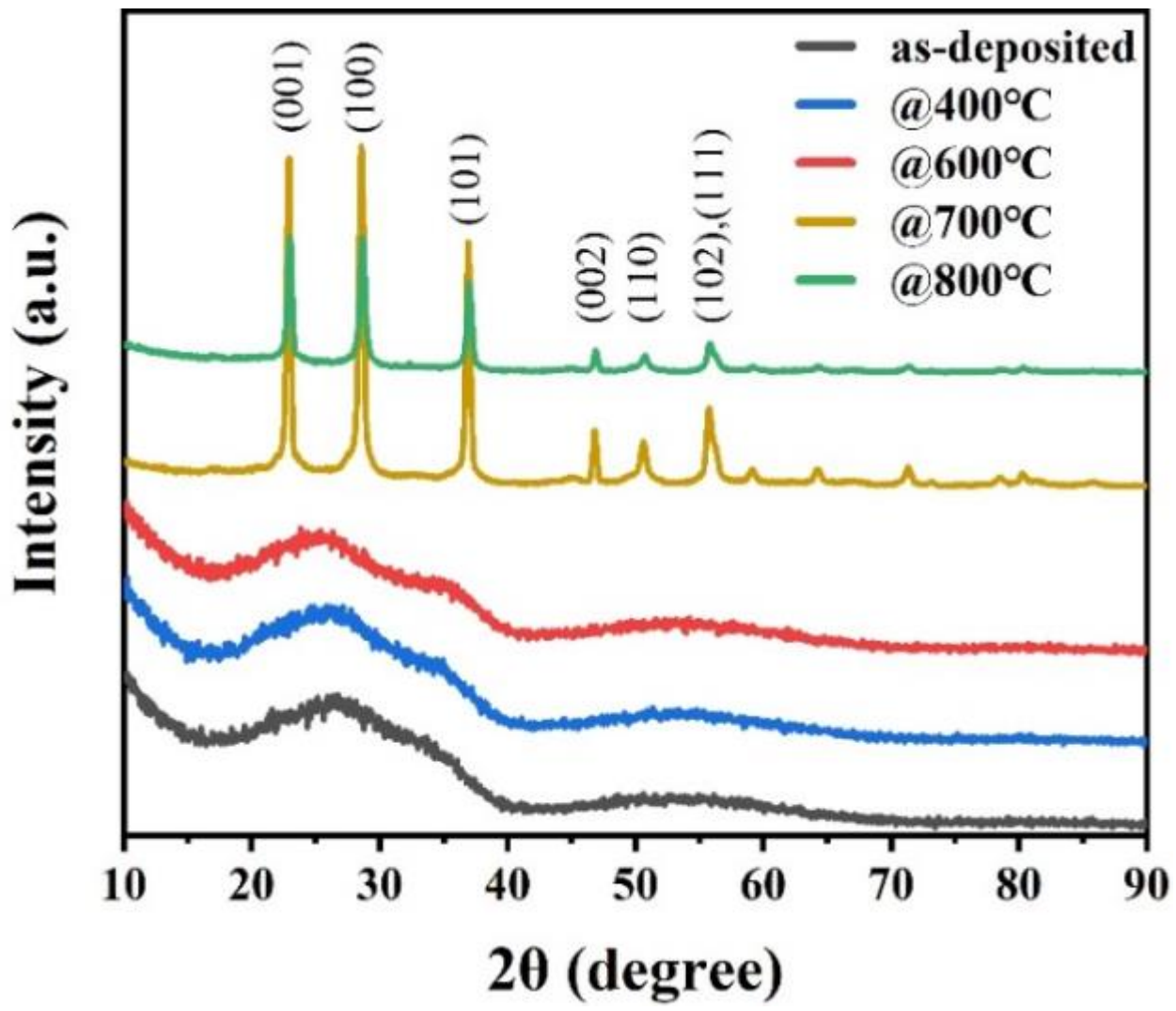

Fig. 2. $Ta_2O_5$ X-ray diffraction measurement results. Below 600 °C the scattering curves (black, blue and red) show broad valleys and humps indicating amorphous phase while at 700 °C and 800 °C sharp peaks arise indicating polycrystalline phase.

Fig. 3(a)–(f) show the two-dimensional surface morphology measured by AFM for different annealing temperature samples. The area of each image is a 5 × 5 μm square, with a measurement accuracy of 0.01 nm. The surface roughness of the fused silica substrate is ~0.65 nm, and the roughness of the as-deposited film decreases. When annealed at 400 °C, the surface roughness decreases to ~0.35 nm. When annealed at 600 °C, the surface roughness slightly increases. The film crystallizes and surface roughness drastically increases to ~3.71 nm at 700 °C. Higher temperature annealing might prompt grain growth and lead to larger grain size. However, it is interesting to notice from Fig. 3(e) and (f) that when annealed at 800 °C the grains become finer compared with 700 °C annealing, which is corroborated by lower XRD peak intensity at 800 °C annealing shown in Fig. 2. This can be explained by stronger pinning effect exerted by Ar bubbles at 800 °C which inhibits grain growth[21]. According to the Zener model, $< R > \propto \frac{r}{f}$, where $< R >$ is the average matrix grain radius, r is the average Ar bubbles radius and f is the Ar bubble area fraction[21,22]. Comparing the STEM images taken at 700 °C and 800 °C as shown in Fig. 4(d, e), we can see clearly that with the coarsening of Ar bubbles at higher temperature the area fraction increases simultaneously. And thus, the average matrix grain radius should decrease with increasing temperature. An alternative model is the pore-grain boundary interaction model[23]. When pores are large and close together the system enters the pore-control regime and the migrating grain boundary is dragged by the pores. This pore-drag mechanism can explain why the larger Ar bubbles at 800 °C are more effective in suppressing grain growth than the smaller bubbles at 700 °C. An increase in grain boundaries with decreasing grain size will usually lead to the rise in RMS surface roughness[24,25]. However, as shown in Fig. 3(g) for 800 °C annealing samples the surface roughness decreases to ~1.70 nm. This can be attributed to increased surface atom mobility which smoothens the surface morphology at higher annealing temperature, which indicates a transition from matrix-relaxation-dominated to surface-diffusion-dominated behavior.

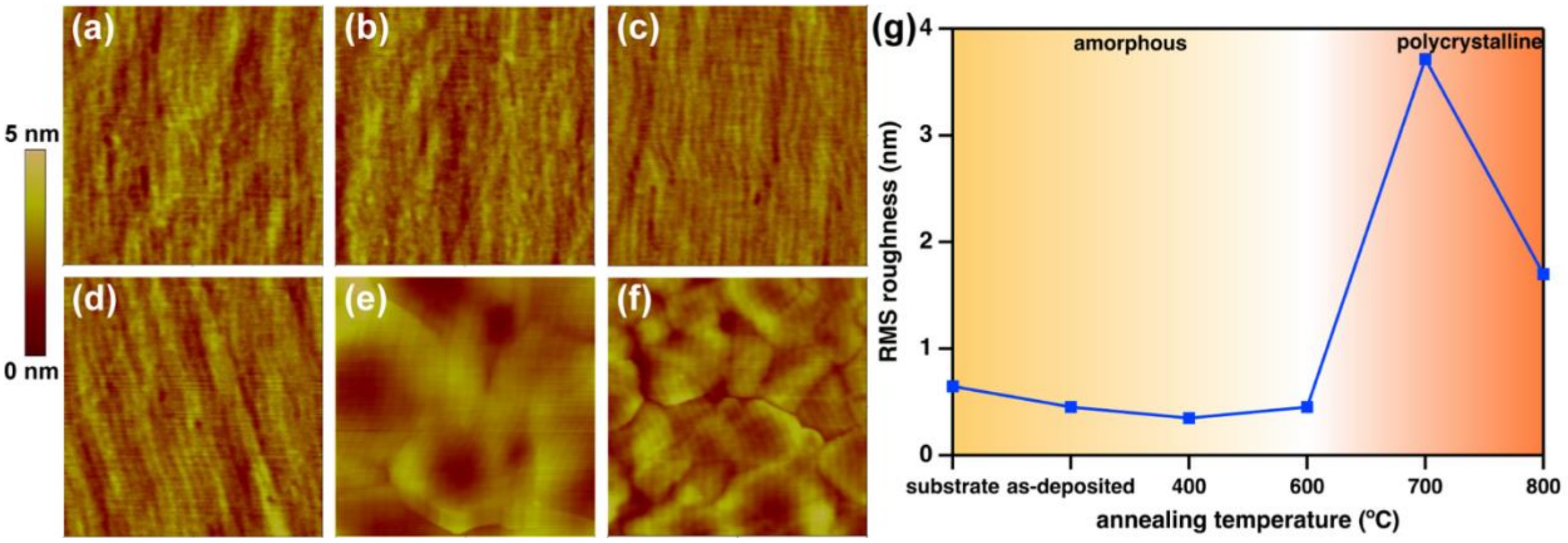

Fig. 3. Surface roughness of $Ta_2O_5$ film at different annealing temperatures. (a) fused silica substrate; (b) as-deposited; (c) annealed at 400 °C; (d) annealed at 600 °C; (e) annealed at 700 °C; (f) annealed at 800 °C; (g) RMS roughness variation with respect to annealing temperature.

STEM experiments were conducted on the samples with different annealing temperatures. Fig. 4(a) shows the unannealed sample, and Fig. 4(b)–(e) correspond to annealed samples at 400 °C, 600 °C, 700 °C, and 800 °C, respectively. The as-deposited sample in Fig. 4(a) already has bubble distributions with a mean diameter of ~2.93 nm. Upon annealing at 400 °C, the mean bubble diameter slightly decreases to ~2.83 nm. This can be understood as the structural relaxation taken place at this temperature leads to densification in the inner region which reduces the average bubble size. At 600 °C, the bubbles coarsen to ~3.10 nm, indicating the onset of thermally activated Ar migration. At this stage, the diffusion of Ar atoms leads to bubble aggregation. A dramatic increase occurs at 700 °C, where the mean diameter jumps to ~4.93 nm as shown in Fig. 4(f), coinciding with the amorphous-to-polycrystalline transition verified by XRD (Fig. 2). Quantitative EDS analysis of the 700 °C annealed sample further confirms that Ar remains localized within these bubbles even after crystallization, showing a higher Ar concentration inside the bubble (~4.29 at. %) compared to the surrounding matrix (~2.39 at. %) (see Supplementary Material, Fig. S2 and Table S1). Further annealing at 800 °C leads to continued coarsening to ~5.46 nm. At this elevated temperature, bubble coarsening is dominated by thermally enhanced diffusion and takes place at grain boundaries[21,26]. As shown in Fig. 4(f), below 700 °C the surface roughness variation follows the same trend as bubble size evolution and therefore bubbles can serve as structural marker. Before crystallization the high-pressure bubbles can exert pressure near the surface region[9] and after crystallization the significantly increased bubbles under the surface can lead to protrusions at the surface and the bubble bursting can lead to surface buckling. However, at 800 °C the bubble size evolution and the RMS roughness decouples. This can be viewed as the competition of two mechanisms: the enhanced surface atoms diffusing tends to smoothen the surface while the bubble coarsening tends to roughen the surface and at this elevated temperature the former mechanism dominates.

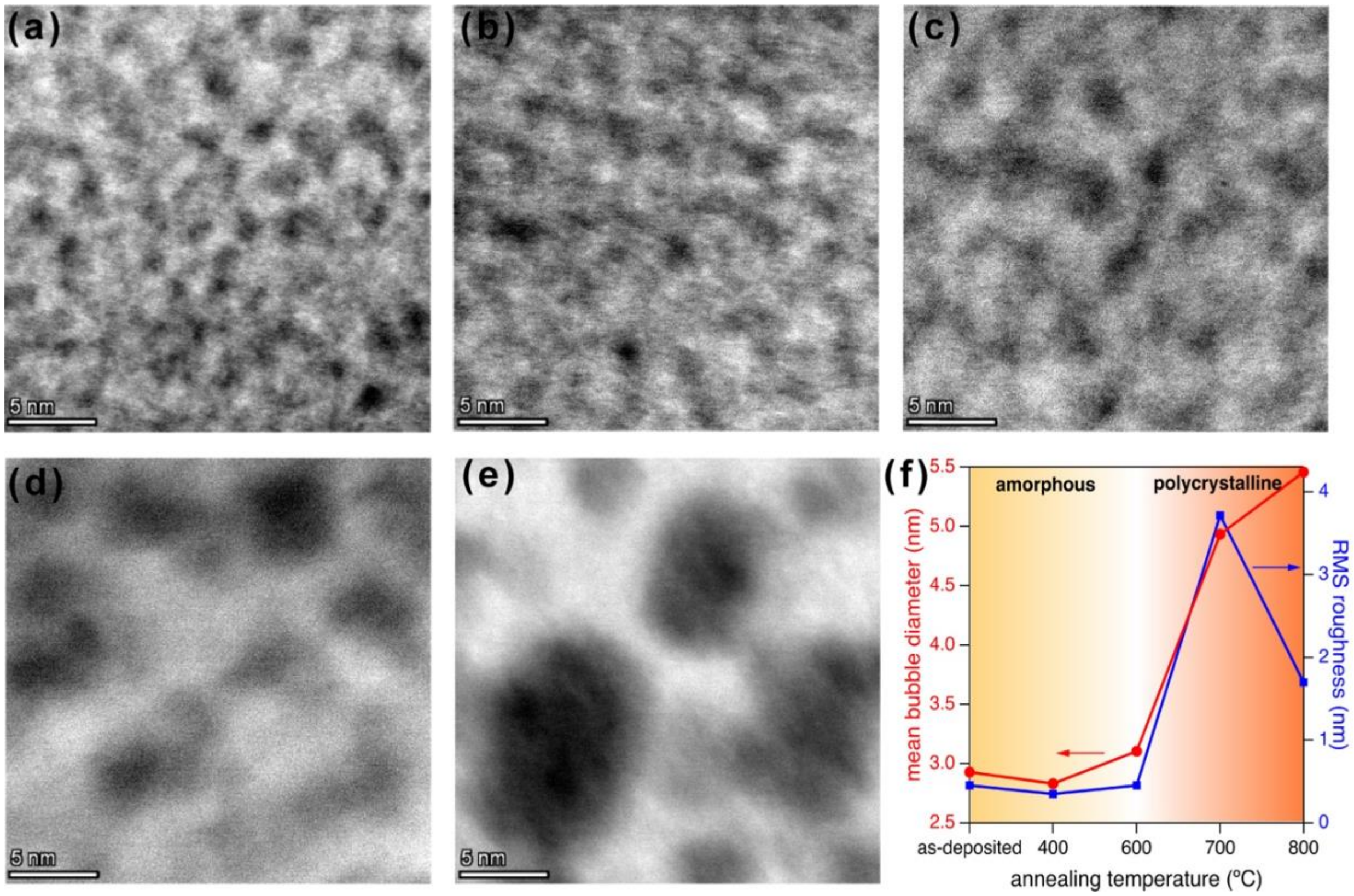

Fig. 4. $Ta_2O_5$ cross-sectional morphology observed under STEM/HAADF. (a) as-deposited sample; (b) annealed at 400 °C; (c) annealed at 600 °C; (d) annealed at 700 °C; (e) annealed at 800 °C; (f) statistical results of average bubble diameter from Fig. 4(a-e) with RMS roughness appended.

### C. Optical evolution and its correlation with structural changes

We further measured transmittance spectra and weak absorption for samples annealed at different temperatures to investigate optical evolution and possible correlation with structural change. As shown in Fig. 5, the spectrum redshift is observed with annealing temperature at 400 and 600 °C, indicating that the optical thickness of films increases after the annealing. From the inset of Fig. 5 we can extract that one of the transmittance peaks located at 559 nm for as-deposited sample while it is positioned at 566.5 nm for 400 °C annealing and 569 nm for 600 °C annealing, which correspond to +1.34% and +1.79% relative shift, respectively. This shift of transmittance peak position is consistent with ~1%-2% film physical thickness increase reported in literature[27]. Thus, the evolution of Ar bubbles plays a minor role in the spectra redshift. After annealing at 700 and 800 °C, transmittance clearly decreases and the spectral curve is distorted, this is primarily induced by scattering from roughened surface and grain boundaries when the annealing temperature reaches the threshold value between 600 °C and 700 °C. Note that as the weak absorption remains at the ppm level as shown in Fig. 6, the decrease in transmittance is not attributed to linear absorption.

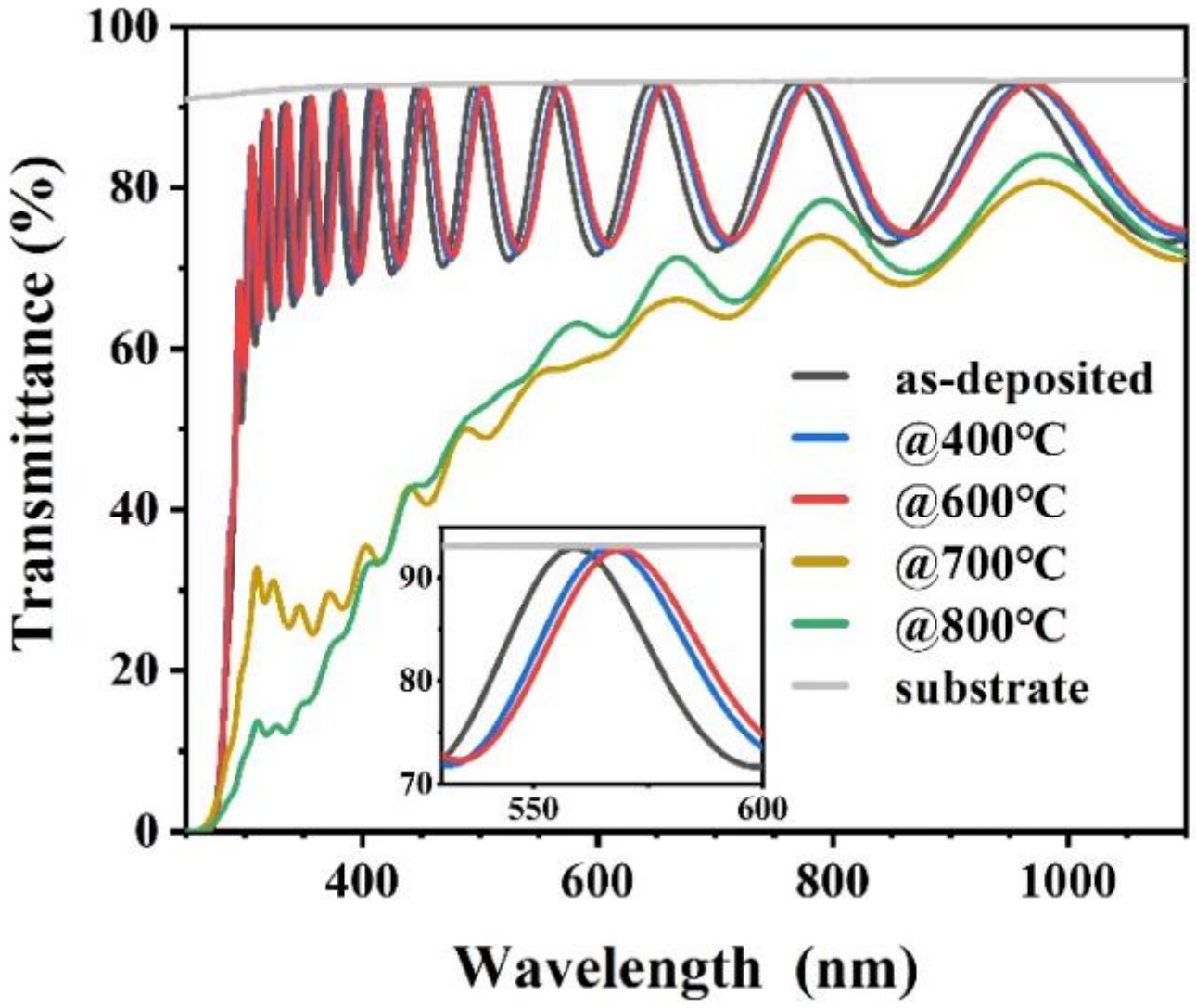

Fig. 5. $Ta_2O_5$ transmittance spectrum at different annealing temperatures. The inset shows the spectrum redshift at annealing temperatures 400 and 600 °C.

The weak absorption of the films at 1064 nm, measured using the surface thermal lens technique, is shown in Fig. 6. The as-deposited samples exhibit a weak absorption of approximately 6 ppm. Upon annealing at 400 °C, the absorption decreases to ~2 ppm and remains at this level after annealing at 600 °C. At 700 and 800 °C, the absorption increases markedly, indicating a degradation of the optical quality at elevated temperatures. In the surface thermal lens measurement, the photothermal signal arises exclusively from the conversion of absorbed photon energy into heat. Under the low-power probing conditions employed here, the dominant heat sources are intrinsic point defects, such as oxygen vacancies and sub-stoichiometric clusters, which absorb 1064 nm photons via single-photon transitions within the bandgap[28]. The photoexcited electrons subsequently relax to their ground states through for example non-radiative electron-phonon coupling, releasing energy as lattice vibrations—macroscopically manifesting as a temperature rise and the surface thermal bump. In stark contrast, Ar bubbles are optically transparent at this photon energy (1.17 eV), possessing no allowed electronic states for single-photon absorption. Moreover, the peak power density in this photothermal setup is orders of magnitude lower than that required to trigger multiphoton ionization within the Ar bubbles.

Consequently, the generation of laser-induced plasma[12] is entirely suppressed under our measurement regime. Therefore, we decouple the role of enlargement of Ar bubbles with weak absorption and attribute the reduction in weak absorption from ~6 ppm (as-deposited) to ~2 ppm (at 400–600 °C) to the thermal annihilation of oxygen vacancies and structural relaxation of the amorphous matrix. The subsequent rise at 700 °C is predominantly governed by crystallization-induced grain boundaries and their associated defect states[29], rather than by the concomitant coarsening of argon bubbles, which serves merely as a structural marker of the phase transition rather than a direct optical loss channel as evidenced by the absorption drop at 800 °C (~3.7 ppm) compared with 700 °C (~8.5 ppm) despite bubble coarsening. This drop can be understood in that 800 °C annealing allows atoms at grain boundaries to overcome potential barriers and find lower energy structures which effectively reduces defect states at grain boundaries[30,31].

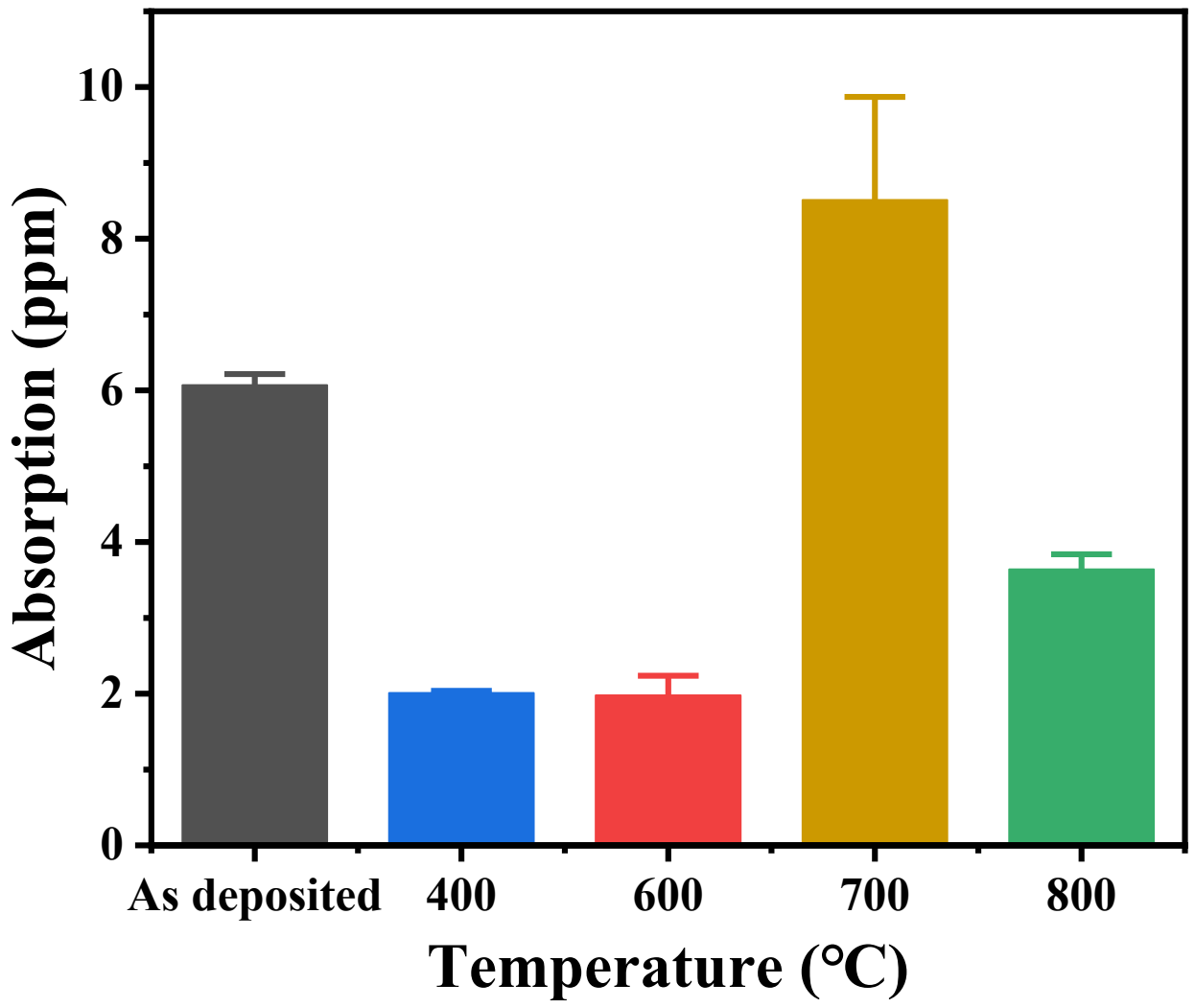

Fig. 6. Weak absorption of $Ta_2O_5$ films at 1064 nm at different annealing temperatures. The absorption decreases from ~6 ppm (as-deposited) to ~2 ppm at both 400 °C and 600 °C, then increases at 700 °C and decreases again at 800 °C.

Based on the above results, annealing-induced structural and optical evolution of IBS-deposited $Ta_2O_5$ films can be categorized into four stages as schematically illustrated in Fig. 7: The first stage is matrix relaxation. At 400 °C, the thermal energy triggers localized structural relaxation and matrix densification but is insufficient to drive extensive Ar atom migration. The dense amorphous network confines the trapped Ar atoms into small, stable nanobubbles. The surface roughness is the lowest and the weak absorption is also low (~2 ppm). The second stage is onset of thermally activated Ar migration. As the annealing temperature rises to 600 °C, thermal activation enables interstitial Ar atoms to migrate through the amorphous network. Aggregation of adjacent Ar atoms causes the bubbles to coarsen slightly. High-pressure bubbles roughen the surface but absorption remains unchanged. The third stage is crystallization and grain boundary absorption. When the temperature reaches 700 °C, a drastic amorphous-to-polycrystalline phase transition occurs. The nucleated crystalline grains rapidly reject insoluble argon atoms[11], which are segregated into the newly formed grain boundaries. The growth of subsurface bubbles exerts strong localized strain, resulting in significant surface roughening. The defect states located at grain boundaries serve as absorption centers and weak absorption rises sharply. The fourth stage is grain boundary relaxation and surface diffusion. At 800 °C, grain boundary relaxation and enhanced surface diffusion reduce the defect density, leading to a decrease in weak absorption and a smoothing of the surface, even though bubbles continue to coarsen. The pinning effect exerted by larger bubbles also leads to smaller grain sizes. The decoupling between bubble size and weak absorption at both 600 °C and 800 °C confirms that linear optical loss is governed by electronic defects rather than argon-filled cavities.

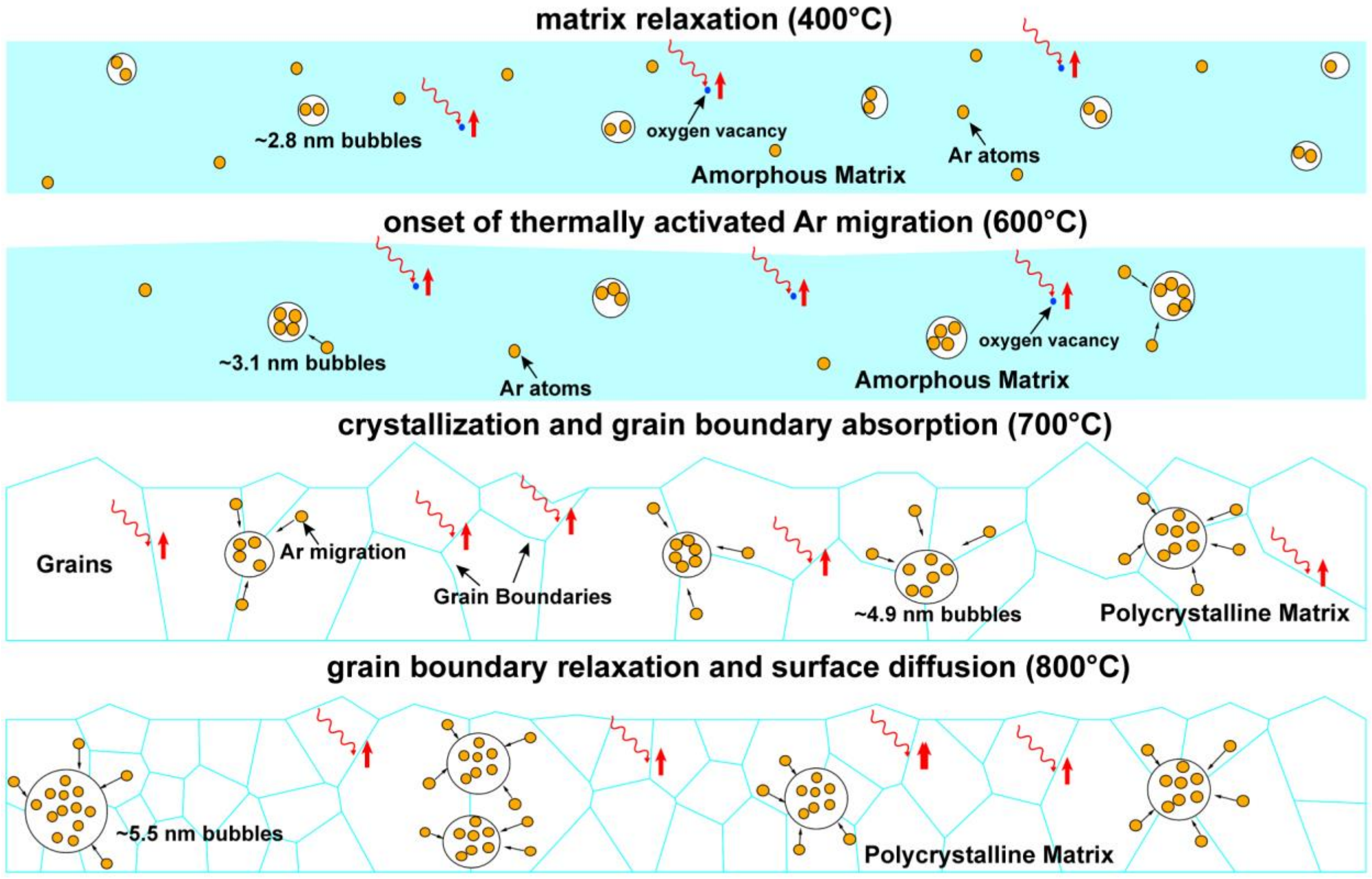

Fig. 7. Schematic diagram illustrating the temperature-dependent structural and optical evolution of IBS-deposited $Ta_2O_5$ thin films: structural relaxation at 400 °C; onset of thermally activated Ar migration at 600 °C; crystallization and grain boundary absorption at 700 °C and grain boundary relaxation and surface diffusion at 800 °C. Red wavy lines and red arrows represent photons absorbed and temperature rise, respectively.

## IV. Conclusion

In this work, we have systematically investigated annealing-induced structural and optical evolution of IBS $Ta_2O_5$ films with particular attention to inherently incorporated Ar bubbles. Ar bubbles in the films serve as structural markers but their evolution trend decouples with surface roughness at 800 °C, indicating a transition from matrix-relaxation-dominated to surface-diffusion-dominated behavior. The weak absorption shows nonmonotonic variation with annealing temperature. It drops from as-deposited to 400 °C and 600 °C and then rises drastically after crystallization at 700 °C and drops again at 800 °C. The decoupling between bubble size and weak absorption at both 600 °C and 800 °C confirms that linear optical loss is governed by electronic defects rather than Ar bubbles. Overall, we categorized the film properties evolution into four stages: matrix relaxation at 400 °C, onset of thermally activated Ar migration at 600 °C, crystallization and grain boundary absorption at 700 °C and grain boundary relaxation and enhanced surface diffusion at 800 °C. Notably, the larger Ar bubbles at 800 °C exert a stronger pinning effect on migrating grain boundaries, resulting in a finer grain size compared to 700 °C. We identified 400 °C as the optimal annealing temperature and 400–600 °C as an effective low-absorption window. Our results provide both fundamental insight and practical guidance for post-deposition thermal treatment of IBS $Ta_2O_5$ films in high-power laser applications.

## Supplementary Material

See supplementary material for data supporting argon aggregation at the hole positions and argon concentration inside bubbles after crystallization.

## Acknowledgments

This work was supported by the National Natural Science Foundation of China (62527901 and 62405340), Shanghai Pujiang Program (24PJD127) and Open Project Program of Wuhan National Laboratory for Optoelectronics (2024WNLOKF008).

## Author Declarations

### Conflict of Interest

The authors have no conflicts to disclose.

### Author Contributions

**Chang Chen**: Investigation; Visualization; Writing – original draft. **Yuan Fang**: Formal Analysis; Visualization; Writing – review & editing. **Yanzhi Wang**: Conceptualization; Supervision. **Jijun Feng**: Supervision. **Yesheng Lu**: Writing – review & editing; Supervision. **Yu Chen**: Supervision. **Ge Zhang**: Investigation. **Jianda Shao**: Funding Acquisition; Supervision.

## Data Availability

The data that support the findings of this study are available from the corresponding author upon reasonable request.